\documentclass[%
 superscriptaddress,
showpacs, preprintnumbers,
 amsmath,
 amssymb,
 aps,
 pra,
showkeys,
onecolumn,
notitlepage,
11pt,
tightenlines
]{revtex4-1}

\usepackage{hyperref}

\usepackage{graphicx}
\usepackage{dcolumn}
\usepackage{bm}

\usepackage[
text={4.6in, 8in},centering,
top=1in,
bottom=1in,
papersize={6.10in,9.25in}
]{geometry}

\let\Oldcite\cite
\renewcommand{\cite}[1]{~\Oldcite{#1}} 
\usepackage{soul}
\usepackage{xcolor}
\definecolor{blueish}{HTML}{08306b}

\usepackage{mathptmx}

\usepackage{titlesec}
\titleformat*{\section}{\large\bfseries}

\titlespacing*{\section}{0pt}{1ex plus 1ex minus .5ex}{0.5ex}
\hypersetup{colorlinks = true,
            linkcolor = blueish,
            urlcolor  = blueish,
            citecolor = blueish,
            anchorcolor = blueish}

\begin{document}

\preprint{\footnotesize To appear in ``\emph{Understanding Crime through Science}" (Springer, 2019)}

\title{Spatial concentration and temporal regularities in crime}
\author{Marcos Oliveira}
\email{moliveira@tuta.io}

\affiliation{GESIS -- Leibniz Institute for the Social Sciences, Cologne, Germany\\}
\author{Ronaldo Menezes}
\affiliation{Department of Computer Science, University of Exeter, UK\\}


\begin{abstract}
\centering
\begin{minipage}{4.5in}
Though crime is linked to different socio-economic factors, it exhibits remarkable regularities regardless of cities' particularities. In this chapter, we consider two fundamental regularities in crime regarding two essential aspects of criminal activity: time and space. For more than one century, we know that (1)~crime occurs unevenly within a city and (2)~crime peaks during specific times of the year. Here we describe the tendency of crime to concentrate spatially and to exhibit temporal regularities. We examine these phenomena from the complex-system perspective of cities, accounting for the possibility of both spatial heterogeneity and non-stationarity in urban phenomena. 
\end{minipage} 
\end{abstract}



\maketitle
\def\pacsname{PACS numberss: }
\def\@keys@name{Keywordsss: }%
\def\frontmatter@PACS@format{%
	\addvspace{21\p@}%
	\footnotesize
	asd
	\adjust@abstractwidth
	as
	\parindent\z@
	\parskip\z@skip
	\samepage
	as
}%


\section{\label{sec:intro}Introduction}
In the last few decades, scientists started to look at cities as evolving systems that exhibit global order built from local-level decisions\cite{Batty1995}. 
Cities bring people together to interact, leading to the emergence of self-organization\cite{Batty1999,Portugali2000,Batty2012,Batty2008}.
Though local-level processes and decisions seem disordered, cities exhibit remarkable regularities that are argued to result from their propensity to expand and to develop\cite{Batty2008,Batty2013b,Bettencourt2007b,Batty2008b,Samaniego2008,DeLong2015,Schiff2015,Bettencourt2016,Youn2016,VanRaan2016,Batty2012}. 
Indeed, evidence of nonlinear growth in urban indicators (e.g.,~wages, serious crime) with city size has motivated scholars to envision cities as complex systems\cite{Batty2013,Bettencourt2010a,Gomez-Lievano2012,Bettencourt2007}. 
From this perspective, researchers have unveiled scaling relationships in criminal activity\cite{Bettencourt2007,Bettencourt2010,Gomez-Lievano2012,Alves2013,Hanley2016} which hints at regularities in crime\cite{Oliveira2015a,Caminha2017,Oliveira2017,Oliveira2018}. 

The existence of scaling suggests general pro\-ces\-ses be\-hind ur\-ban development\cite{Bettencourt2013hyp}. 
It indicates a general mechanism underlying urbanization and implies that regularities exist in cities regardless of their idiosyncrasies. 
The study of urban scaling has provided the means to understand urban growth and its impact on indicators such as employment, patent, wage, and crime\cite{Bettencourt2007,Bettencourt2010,Bettencourt2013,Gomez-Lievano2016}. Most of these analyses, however, neglect details of the indicators such as spatial distribution across the city, probably because of the lack of fine-grained data\cite{Oliveira2017}. In the case of crime, researchers have taken advantage of the availability of data to study the phenomenon\cite{White2015,PachecoOM17,Kadar2018,DaCunha2018} and to describe its regularities\cite{Davies2015,Oliveira2015a,Alves2015,Caminha2017,Oliveira2017,Oliveira2018}.

Evidence of regularities in crime traces back to the nineteenth century. 
Almost simultaneously, Adolphe Quetelet and Andr\'{e}-Michel Guerry were the first to describe regularities in criminal activity\cite{friendly2007,quetelet1842treatise,BalbiGuerry:1829,quetelet1833recherches}. 
For almost two centuries, crime in cities has been known to exhibit seasonality and to be unevenly distributed\cite{BalbiGuerry:1829,quetelet1833recherches}. 
With these findings, Quetelet and Guerry pioneered the viewpoint of physical laws governing human populations\cite{friendly2007}. 
Though distinct socio-economic factors influence crime, remarkable regularities exist in its dynamics. 
In this chapter, we discuss regularities in crime concerning its spatial distribution and its temporal dynamics. 

In the case of spatial regularities, researchers have found that criminal events tend to cluster spatially\cite{Weisburd2015}. Offenses concentrate in such a way that most of the occurrences happen at very few places. 
This concentration has been confirmed regardless of spatial granularity levels in a myriad of studies. The overwhelming amount of evidence makes us expect that, in any city, some areas will have disproportionately more crime than others\cite{Weisburd2015}. 
Such ubiquity has brought to this observation the status of a law, namely the \emph{law of crime concentration}, which states that a small number of micro-geographic units account for most of the offenses in a neighborhood or city\cite{Weisburd2015}. In Section~\ref{sec:concentration}, we discuss the characteristics of crime concentration in cities and its relationship with the type of crime and city size.

In addition to displaying spatial regularities, crime is known to depend on time constraints and to exhibit temporal regularities\cite{Cheatwood2009}. Much research has been devoted to describing periodic changes in crime rates such as annually and weekly. Adolphe Quetelet linked weather variations to aggression to explain criminal seasonality\cite{quetelet1842treatise,Cheatwood2009}. 
From his perspective, the heat would cause the stress needed to turn people more likely to offend. His so-called ``thermic law" of crime, however, has been since replaced by an understanding of indirect relevant effects\cite{Harries1984,Cheatwood2009}. 
For instance, fluctuations in social dynamics may affect three requirements for crime: {\em offender}, {\em target}, and {\em opportunity}\cite{Cohen1979,Cheatwood2009}. 
In Section~\ref{sec:rhythms}, we discuss the temporal regularities in crime when we account for both spatial heterogeneity and non-stationarity in cities.

The regularities as mentioned earlier have a direct impact on how we see cities. With the their characterization, we provide a pathway for researchers to create realistic models of crime and present the grounds to understand the impact of local activities on global patterns of cities. 
In the last section of this chapter, we discuss such aspects and the implication to policymaking. 

\section{Spatial concentration}
\label{sec:concentration}
\begin{figure}[t!]
\begin{minipage}{4.6in}
    \centering
    \includegraphics[width=4.6in]{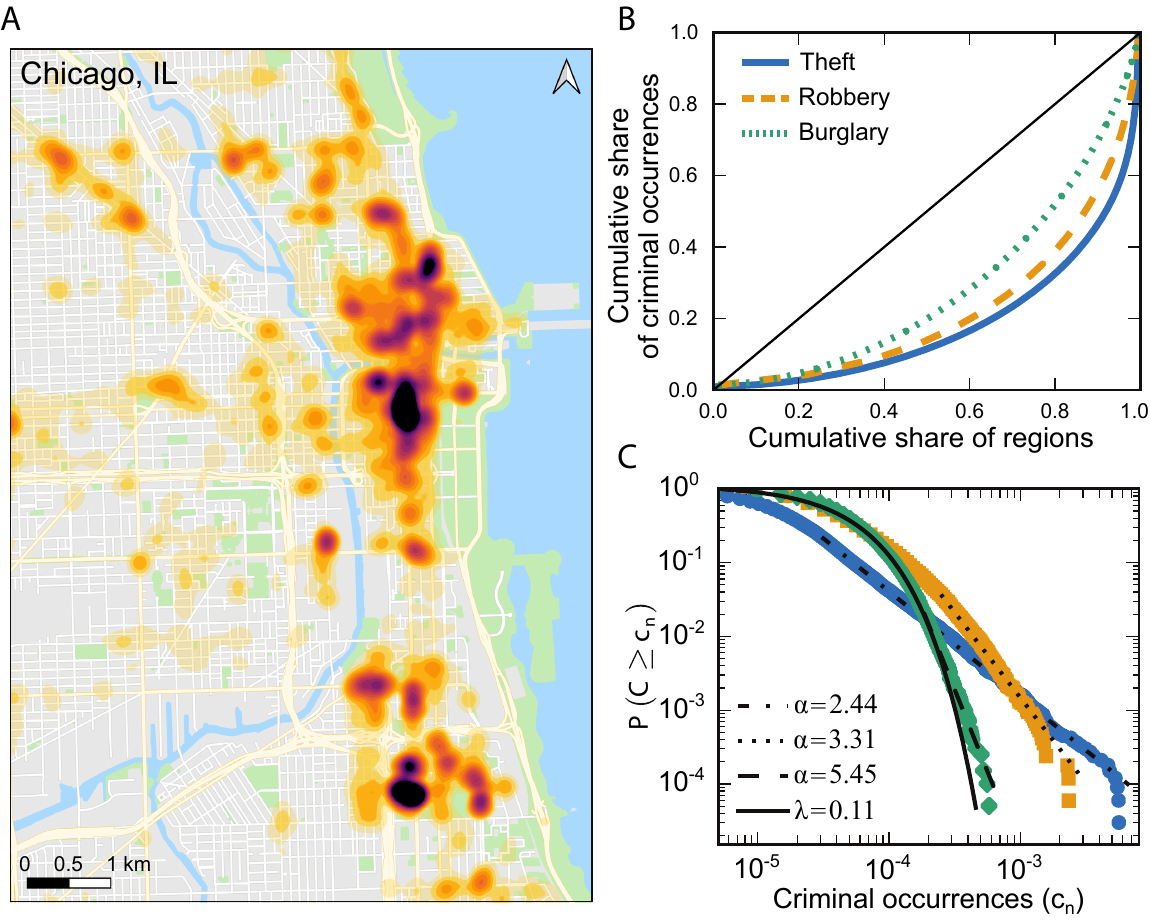}
    \caption{\textbf{Crime clusters spatially with a concentration level that depends on the offense type.}
    For example, (\textbf{A}) depicts the spatial distribution of theft in Chicago,~IL. In the figure, darker colors mean higher crime rates. This map illustrates the tendency of crime to cluster in specific regions of Chicago. This concentration depends on the type of offense. Each curve in (\textbf{B}) shows the relationship between the cumulative crime share and the cumulative share of the corresponding number of regions ordered by crime counts (i.e., Lorenz curves). Theft concentrates more than robbery, and robbery more than burglary. This spatial concentration of crime can be describe with (\textbf{C}) a power-law distribution $p(x) \propto x^{-\alpha}$ where the exponent $\alpha$ relates to the type of crime. Figures adapted from\cite{Oliveira2017}. 
    }
    \label{fig:fig1}
	\end{minipage}
\end{figure}

The spatial concentration of crime has been confirmed in different cities using various spatial aggregation units including street and area level (e.g.,~census tracts, street segments)\cite{EckE.2007,Johnson2010,Braga2010,PrietoCuriel2017}. 
For instance, Fig.~\ref{fig:fig1}A depicts the spatial distribution of thefts in Chicago, IL. In this example, we can see some regions of the city having disproportionately more criminal activity than others. 
We want to characterize the phenomenon over different cities to further study this spatial regularity. For this, first, we need a general approach to analyze the city. With a general method, we avoid biases due to cities' particularities (e.g., blocks size, length of the street segments)\cite{Oliveira2017}. 
We can achieve this by building spatial aggregation units based on the population distribution across the city. Specifically, we divide the area of a city into regions with equal population size  (i.e., number of residents), then we aggregate offenses that happened within the same regions.

\begin{figure}[t!]
  \begin{minipage}{4.6in}
    \includegraphics[width=4.6in]{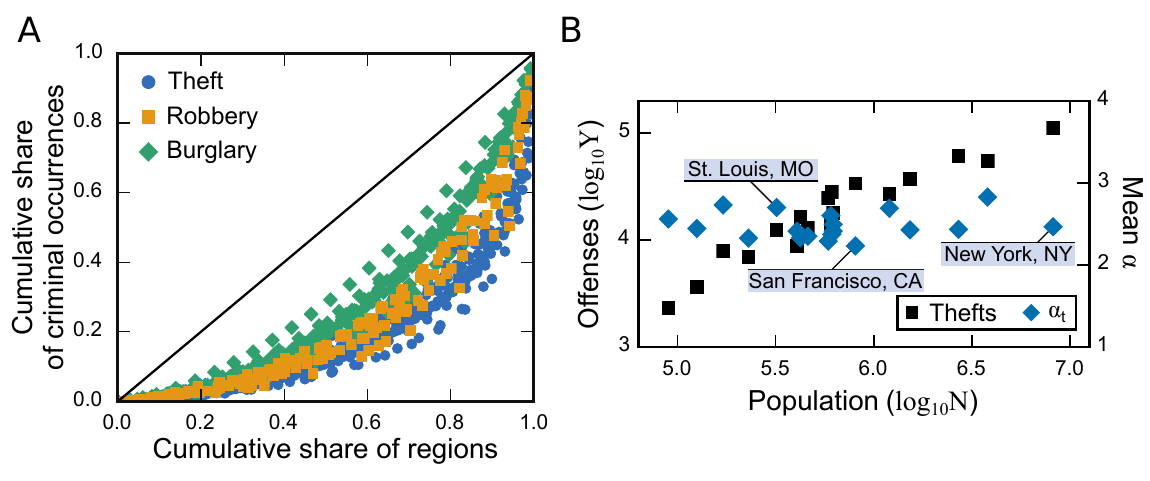}
    \caption{\textbf{Crime concentration is  scale-invariant but depends on crime type.} 
    With data from 25 cities, the Lorenz curves in (\textbf{A}) demonstrate the dependence between the type of crime and concentration level. This concentration, however, is independent of city size. (\textbf{B}) depicts a comparison between concentration exponent for theft, population size, and the number of thefts. 
In this double-y-axes plot, squares represent the number of crime in a city during one year, and diamonds stand for the average power-law exponent for a city. Though population size has a clear relationship with the number of offenses, it fails to have an association with $\alpha$. This independence is confirmed using the Hoeffding's test of independence. Figures adapted from\cite{Oliveira2017}.
    }
    \label{fig:fig2}
  \end{minipage}
\end{figure}

With these counts, we can now analyze the empirical distribution of crime in cities. Fig.~\ref{fig:fig1}B shows the Lorenz curves of the crime distribution in Chicago for burglaries, robberies, and thefts. 
The Lorenz curves help us to assess the concentration of crime.
The curves in Fig.~\ref{fig:fig1}B indicate the tendency of crime to concentrate spatially regardless of the offense type. However, note that the \textit{level} of concentration seems to depend on the offense type. 
A similar tendency also occurs in $25$ other locations from the United States and the United Kingdom:~theft concentrates more than robbery, and robbery more than burglary (see Fig.~\ref{fig:fig2}A).

To describe these quantities statistically, we can fit the distribution of crime with different distributions and then compare them using the likelihood ratio test\cite{Oliveira2017}. In the majority of the $25$ cities, the probability distribution of crime across a city can be described by a power law
$p(x) \propto x^{-\alpha}$
where the exponent $\alpha$ relates to the type of crime. The different types of crime present distinct levels of concentration which manifests on the range of the power-law exponent: for thefts, $\alpha_t$ is between $2.1$ and $3.0$; whereas the exponents for burglaries $\alpha_b$ and robberies $\alpha_r$ vary in wider ranges with $\alpha_r$ within $2.4$ and $4.1$, and $\alpha_b$ between $2.9$ and $6.0$. The different dynamics of each type of crime might be the cause of the distinct exponent intervals and thus the level of concentration.
Note that, in some cases that $\alpha$ exhibits large values (e.g., burglary), the exponential and the power-law distributions are both good descriptions of the data\cite{Oliveira2017}.

We also need to study the stability of the concentration, since such a high level of concentration could indicate a somehow static city\cite{Oliveira2017}.
To examine the stability, we can measure the uncertainty of the positions in the rank of criminal spots over time. 
Precisely, we calculate the Shannon entropy  ${H}_{r_t}(i)$ of each position $i$ in the criminal ranks of regions $r_t$ which are created using the number of offenses weekly aggregated\cite{Oliveira2017}. 
Using data from $20$ cities of the U.S., the mean normalized entropy (i.e., the sample mean over the positions in the rank) is around $0.95$, indicating that criminal spots are likely to vary across regions over time. 
The hottest spots (i.e., the first positions in the rank), however, present distinct dynamics with the entropy ${H}_{r_t}(i)$ increasing quickly with the position $i$. This result means that the regions with the most criminal activity tend to be the same ones. 
Similarly, the places with the least criminal activity (i.e., the last positions in the rank) are usually the same regions.
This analysis also reveals that the rank of thefts presents lower entropy in the first rank positions in comparison to the other types of crime\cite{Oliveira2017}.
In other words, we have more certainty about the whereabouts of the hottest spots of theft than the hottest spots of robbery and burglary. 

These findings indicate the level of spatial crime concentration as a regularity that occurs regardless of idiosyncrasies of the city. In particular, this is an intriguing finding because of the allometric scaling of crime in cities. 
Fig.~\ref{fig:fig2}B shows the relationship between $\alpha_t$ of some U.S. cities and their population size. On the one hand, crime numbers relate to city size; on the other hand, crime concentration (i.e., the exponent) seems to be independent of city size. 
We can study the statistical dependence of this relationship using the Hoeffding's test of independence\cite{nsm}. 
Specifically, we can test the relationship between the population size of the cities and the average power-law exponent $\alpha$. 
With data from the considered U.S. cities, we could not reject the hypothesis that the size of the city and the level of crime concentration are independent with the 95\% confidence\cite{Oliveira2017}. This result suggests crime concentration as an attribute of criminal phenomena which occurs regardless of the population size of the city.

\section{Criminal rhythms}
\label{sec:rhythms}
Much effort has been devoted to studying temporal regularities in crime since Quetelet's seminal work on crime seasonality\cite{BAUMER1996} . Researchers have confirmed annual seasonality in several cities and studied other regularities related to different temporal conditionals such as the day of the week, the hour of the day, and the presence of holidays\cite{BAUMER1996}.  
Most of the studies in the literature, however, assume a  temporal regularity of crime activity limited within fixed regional localities\cite{Oliveira2018}.  
From the perspective of cities as complex systems, this assumption neglects the continuous process of organization in cities\cite{Batty1995,Batty2008}.
Such stationary assumption implies that the urban dynamics in all regions across the city remain constant over time\cite{Oliveira2018}.

To study crime from the viewpoint of dynamic cities, we have to consider the possibility of non-stationarity and spatial heterogeneity. 
For this task, we can use wavelet analysis to track the periodicity of a criminal time series. With wavelet analysis, we can evaluate the statistical significance of a component (e.g., annual, biannual) in a time series over time. If we account the spatial heterogeneity, we are also able to describe how the rhythms of crime are distributed across the city and how they vary over time.  

\begin{figure}[t!]
  \begin{minipage}{4.6in}
    \includegraphics[width=4.6in]{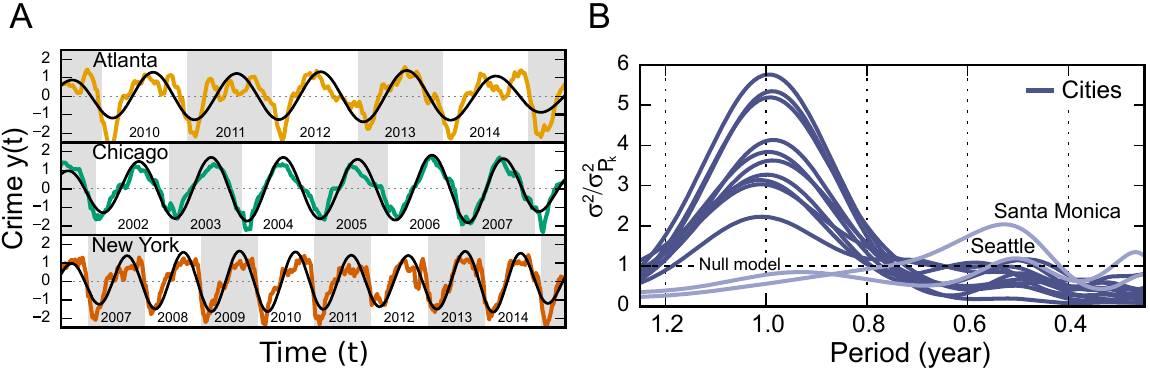}
    \caption{\textbf{Crime exhibits a circannual cycle at the city level.} In (\textbf{A}), the time series of theft $y(t)$ for selected cities indicate the rhythms of crime. With crime data from 12 cities, this cycle is confirmed using (\textbf{B}) the global wavelet spectrum of the wavelet transform of $y(t)$ and comparing against the null model $P_k$ generated from autocorrelated random noise (dashed line); two cities, however, failed to exhibit this periodicity. This striking temporal regularity is seen in (\textbf{A}) the waves (black curves) that are reconstructed using only the circannual band. Figures adapted from\cite{Oliveira2018}.
    }
    \label{fig:fig3}
	\end{minipage}
\end{figure}

In our case, a time series is a discrete sequence $\smash Y = \left\{y\left(1\right), \hdots, y\left(N\right)\right\}$ with observations of uniform time step $\delta t$ in which $y(t)$ represents the processed number of thefts (i.e., without long-term trends) at the week $t$. Here we denote $Y^c$ as the city-level time series of city $c$, whereas $Y^c_i$ represents the time series of the region $i$ in the city $c$. For instance, Fig.~\ref{fig:fig3}A depicts $Y^c$ of selected cities from the United States. 

First we study the temporal regularities (i.e., periodicity) in $Y^c$ using its wavelet transform $\smash{{\rm W}_{Y^c}(s,n)}$.
This transform gives us the contribution of a periodicity (scale) $s$ at different moments $n$ in the time series. The \emph{local wavelet spectrum}, defined as $\smash{|{\rm W}_{Y^c}(s,n)|^2}$, enables us to identify temporal regularities in a series. 
For this, we average the spectrum across either time $n$ or period $s$. 
To identify cycles in the entire series, we average $\smash{{\rm W}_{Y^c}\left(s,n\right)}$ over $n$, known as the \emph{global wavelet spectrum} and denoted as $\smash{\overline{{\rm W}}^2\left(s\right)}$, which provides us an averaged picture of the periods in the time series\cite{Percival1995}. Fig.~\ref{fig:fig3}B shows the global spectrum of some cities in the United States. The peak around $1.0$ indicates the existence of a statistically significant annual periodicity of crime. This analysis confirms previously well-documented evidence that crime exhibits circannual cycles\cite{BAUMER1996}. 
 
\begin{figure}[t!]
  \begin{minipage}{4.6in}
    \centering
    \includegraphics[width=4.6in]{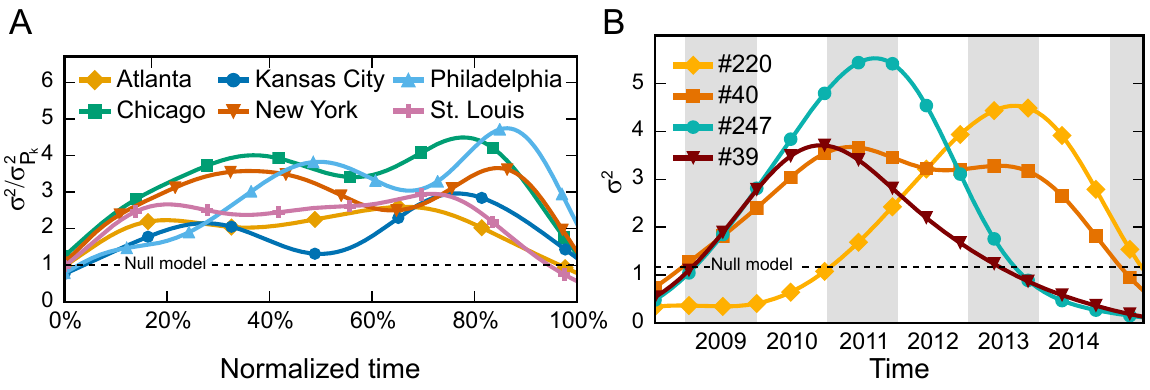}
    \caption{\textbf{The circannual cycle is stationary at the city level but non-stationary at the local level.}
    In (\textbf{A}), the scale-averaged wavelet power of the wavelet transform of the city-level time series with respect to the circannual band shows that the city-level time series are stationary over time (selected cities in the figure). At the local level, however, the time series are non-stationary. For instance, (\textbf{B}) depicts the scale-averaged wavelet power of four regions in Chicago. These curves exemplify local-level time series changing their circannual periodicity over time. Figures adapted from\cite{Oliveira2018}.}
    \label{fig:fig4}
	\end{minipage}
\end{figure}

Note that $\smash{\overline{{\rm W}}^2\left(s\right)}$ is an average over time $n$ and thus neglects any temporal dynamics such as non-stationarity. To test the stationarity in these regularities, we average $|{\rm W}_{Y^c}(s,n)|^2$ over scale $s$, yielding the \emph{scale-averaged wavelet power}. It enables us to analyze the temporal evolution of a periodic signal in terms of a given band $b=\left(j_1, j_2\right)$. In our case, we want to examine the circannual stationarity, so we evaluate the scale-averaged wavelet power with $j_1=0.8$ and $j_2=1.1$ for each city and test this periodicity against a null model. Fig.~\ref{fig:fig4}A shows the scale-averaged wavelet power for some cities. Most of them exhibit stationary time series; that is, the circannual component is present throughout the series. 
In fact, this stationarity is already evident in Fig.~\ref{fig:fig3}A where  the circannual component stays the same over time. 

At such a city level, crime has a striking temporal regularity that offers the impression of cities as stationary objects. This analysis, however, neglects spatial heterogeneity. The continuous organization process in cities suggests local-level dynamics changing across the city.
We expect variations that result from changes in human dynamics at the local and global levels of the city (e.g., an influx of new residents, closing establishments, new subway stations). To have a local-level view of cities, we can examine the time series from smaller spatial aggregation units. Similarly to Section~\ref{sec:concentration}, we divide each city $c$ into regions of similar population size and build the time series $Y^c_i$ for each region $i$. 

Now we can examine the stationarity of crime at the local level. Fig.~\ref{fig:fig4}B shows the scale-averaged power of three regions in Chicago regarding the circannual component. Though they all exhibited a circannual cycle at some interval, they have distinct dynamics. 
For example, region {\tt \#40} presents stationarity while region {\tt \#247} loses its circannual periodicity in 2014. 
Though at the city level crime exhibits stationarity, circannual rhythms change over time at the local level. The aggregated data hide local dynamics. In fact, other periodicities take place at lower spatial granularity such as biannual and triannual but disappear at city level\cite{Oliveira2018}.

With all the time series $Y^c_i$ of a city, now we can examine crime from bottom to up. 
We can study the non-stationarity in cities at the local level from a holistic perspective. For this, we count the number of regions that significantly show the circannual period at each time step. Specifically, the \textit{composed scale-averaged power} $C^b(t)$ is defined as the number of regions that exhibit a statistically significant band $b=(j_1, j_2)$ at the time step~$t$\cite{Oliveira2018}. 
With data from $12$ cities in the U.S., we can show that $C^b(t)$ exhibits a typical value without much variability over time;~that is, cities stay with a similar number of regions with 1-year cycle over time. 
This result is intriguing: though the time series are non-stationarity, the number of time series with a circannual period remains somewhat the same throughout the series. 

Yet, note that $C^b(t)$ considers only the \textit{number} of regions and neglects the regions themselves.  
We do not know if the number of regions stays the same because of the \textit{same set} of regions.  
To understand better $C^b(t)$, we are also interested in the amount of time $\Delta t$ that a region exhibits a significant circannual periodicity. Precisely, we count the number of weeks that each region keeps the circannual band significant continuously. 
From all the considered cities, $\Delta t$  decays much earlier than the total time of the criminal series. 
That is, in general, the amount of time that a region has a circannual cycle is shorter than the entire time series. 
This result coupled with the form of $C^b(t)$ implies waves of crime traveling across the city---a finding that agrees with the notion of cities continuously changing over time. 

\section{Discussion}
Cities are evolving systems that exhibit global phenomena emerging from local-level actions, presenting messy but ordered patterns at different levels\cite{Batty2008,Oliveira2018}. From this perspective, we expect to find regularities in crime which transcend cultural and socio-economic particularities of cities. With the description of these regularities, we move towards better models to understand cities. In this chapter, we considered regularities present in two fundamental aspects of crime: space and time.  

The spatial concentration of crime is not only ubiquitous in cities but also independent of the city size. The concentration level, however, depends on the type of crime---perhaps because of different kinds of crime exhibit particular dynamics. Such concentration coexists with the continuous displacement of crime spots in cities. That is, crime constantly flows across the city while maintaining the system organization in a way that its dynamics and regularities appear to be scale-invariant. These features suggest an understanding of crime from a complex-system perspective. 

The well-documented seasonality in crime has a different picture when we admit cities as dynamic and continually organizing processes. When we account for spatial heterogeneity, we find that the circannual cycles of crime are unevenly distributed across the city. When we also consider non-stationarity, we observe features that agree with complex cities. On the one hand, the seasonality is stationary at the city level, but on the other hand, the criminal waves are non-stationary at the local level---they travel across the city. These findings support analyzing crime under the perspective of dynamic cities.

The complex-system perspective of crime asks for tools and analyses that cover the system as a whole in order to understand crime dynamics. Indeed, the connectedness of the city rejects studies focusing on the hotspots of crime and neglecting the ``cold" areas. In this scenario, network-based approaches might be essential to handle such mezzo level\cite{White2015,Oliveira2015a,Davies2015,DaCunha2018}. 

The regularities discussed in this chapter  have impacts on policy-making. The independence of crime concentration with city size implies that high crime regions are expected to exist as the city grows, urging for proper government policies. 
Still, cities continuously change over time; thus policy-making needs evolving approaches and a constant assessment of the city. In this regard, policy-makers need tools and up-to-date data to assess the changes happening in cities. With the proper tools, we can learn more about cities and help to improve them.

\bibliographystyle{naturemag} 

\end{document}